\documentclass{bmvc2k}

\title{A2V: A Semi-Supervised Domain\\
Adaptation Framework for Brain Vessel\\
Segmentation via Two-Phase Training Angiography-to-Venography Translation}

\addauthor{Francesco Galati}{galati@eurecom.fr}{1}
\addauthor{Daniele Falcetta}{}{1}
\addauthor{Rosa Cortese}{}{2}
\addauthor{Barbara Casolla}{}{3}
\addauthor{Ferran Prados}{}{4,5}
\addauthor{Ninon Burgos}{}{6}
\addauthor{Maria A. Zuluaga}{}{1,7}

\addinstitution{
 EURECOM, Biot, France
}
\addinstitution{
 University of Siena, Italy
}

\addinstitution{
 Centre Hospitalier Universitaire de 
 \\ Nice, France 
}

\addinstitution{
Centre for Medical Image Computing\\
University College London, UK
}
\addinstitution{
 e-health Center \\
 Universitat Oberta de Catalunya \\
 Barcelona, Spain
}
\addinstitution{
Sorbonne Université, Paris Brain\\
Institute, CNRS, Inria, 
Inserm, \\
AP-HP, Paris, France
}
\addinstitution{
School of Biomedical Eng. \& Life \\
Sciences, King's College London, UK
}

\runninghead{Galati et al.}{A Domain Adaptation Framework for Vessel Segmentation }

\usepackage{amssymb}

\usepackage{tabularx}
\usepackage{booktabs}
\usepackage{multicol, multirow}
\newcolumntype{Y}{>{\centering\arraybackslash}X}
\newcolumntype{Z}{>{\hsize=.6\hsize}X}

\usepackage{tcolorbox}

\begin{document}

\maketitle

\begin{abstract}
We present a semi-supervised domain adaptation framework for brain vessel segmentation from different image modalities. Existing state-of-the-art methods focus on a single modality, despite the wide range of available cerebrovascular imaging techniques. This can lead to significant distribution shifts that negatively impact the generalization across modalities. By relying on annotated angiographies and a limited number of annotated venographies, our framework accomplishes image-to-image translation and semantic segmentation, leveraging a disentangled and semantically rich latent space to represent heterogeneous data and perform image-level adaptation from source to target domains. Moreover, we reduce the typical complexity of cycle-based architectures and minimize the use of adversarial training, which allows us to build an efficient and intuitive model with stable training. We evaluate our method on magnetic resonance angiographies and venographies. 
While achieving state-of-the-art performance in the source domain, our method attains a Dice score coefficient in the target domain that is only $8.9\%$ lower, highlighting its promising potential for robust cerebrovascular image segmentation across different modalities.

\end{abstract}

\section{Introduction}
\label{Introduction}

\begin{figure}
\centering
\includegraphics[width=\textwidth]{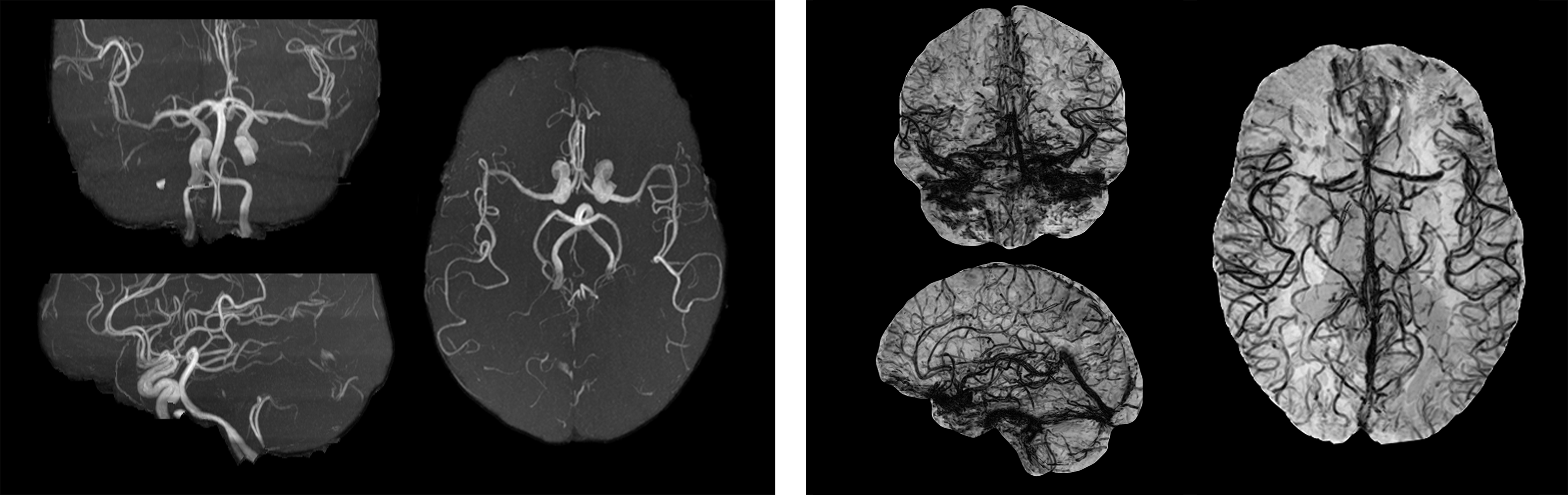}
\caption{Maximum intensity projection (MIP) of a magnetic resonance angiography (left) and minimum intensity projection (mIP) of a magnetic resonance venography (right) along the three spatial axes.} \label{Fig:ex_TOF_and_SWI}
\end{figure}

The accurate segmentation of the cerebrovascular tree is critical to multiple diagnostic and therapeutic procedures and to the study of the brain's health and associated pathologies. Due to the intricate morphology of the cerebrovascular tree~\cite{Helthuis2019}, different clinical applications require different imaging techniques to visualize and analyze the brain vessels. Besides the differences in the image appearance and intensity distributions, image modalities may also vary in the type of vessels they target, i.e., either arteries or veins (see example in Figure~\ref{Fig:ex_TOF_and_SWI}). This diversity results in a large distribution shift that causes automatic segmentation models to fail when used across different image modalities. Given the variety of imaging techniques to assess the brain vasculature~\cite{Lin2018}, the lack of cross-modality generalization is particularly limiting. Training, deploying, and maintaining a segmentation model per image modality is an expensive and highly demanding task.  

Domain adaptation (DA) has been an active research field to tackle the problem of distribution shift, also known as domain shift, by finding a mapping from the source data distribution to the target distribution with little to no labels available in the target domain~\cite{Guan2022}. Recent studies have shown promising results for different organs~\cite{Guan2022},
among which the brain and its main substructures~\cite{SynthSeg,Kamnitsas2017a,Yao2022}. Nevertheless, no current work has focused specifically on the brain vessels. This can be explained by two factors. First, brain vessels are relatively small objects within a large image volume~\cite{Dang2022} and the preservation of fine details is still a challenging task for state-of-the-art DA methods. As a consequence, small objects are often merged with the background, thus impeding their segmentation~\cite{Tsai2018}. Second, while there is a correlation between the morphology of arteries and veins, they occupy different positions in the brain, and arteries generally have larger sizes. This results in a domain gap that encompasses variations in the overall image appearance, but also includes dissimilarities in the structure and spatial arrangement of cerebral vasculature between the source and target domains. As a result, there are currently no methods that can seamlessly be used to segment both the arteries and the veins.

Given that brain arterial segmentation is a simpler and more widely explored task~\cite{Moccia2018}, we aim at segmenting brain veins by relying 
primarily on annotated data from angiography images. 
%
To this end, we develop a semi-supervised domain adaptation framework that accomplishes both image-to-image translation and semantic segmentation of the cerebrovascular tree, relying on a disentangled and semantically rich latent space to represent heterogeneous volumetric data and to perform image-level adaptation from source to target domain, where vessels have different properties. The ability to disentangle volume-related and vessel-related image properties enables our method to generate a translation that transforms veins into arteries only in appearance, while preserving crucial spatial information such as shape and location. This bridges the gap between the two domains, which in turn enables easier generalization from labeled source data to unlabeled target data and thereby facilitates the segmentation of the vessels in both domains. Additionally, we reduce the typical complexity of cycle-based architectures~\cite{CycleGAN} by decreasing the number of components from multiple to three: one generator, one discriminator, and one encoder. This reduction leads to more efficient utilization of computational resources and results in a simpler model, enabling faster experimentation at training and easier deployment of the model in real-world scenarios. We assess our method in the segmentation of vessels from magnetic resonance (MR) angiography and venography images, while mainly relying on annotated angiographies. Our results are particularly promising for the application of DA methodologies to the less-explored area of brain vein segmentation.

\section{Related Work}
\label{Related Work}
{\color{bmv@sectioncolor}\textbf{Multi-modal brain vessel segmentation.}}
Although 3D brain vessel segmentation has been extensively studied~\cite{Moccia2018}, few works have focused on the problem of cross-modality generalization. In \cite{passat2007}, the authors use morphological operators on paired T1-weighted MR and time-of-flight magnetic resonance angiography (MRA) sequences, whereas \cite{zuluaga2015} presents a tensor voting strategy using paired computed tomography angiography and 3D phase-contrast MR images. Although these works use the same algorithm to segment arteries from different modalities, each modality requires a different initialization process or modality-specific parameter tuning. Most recently, \cite{Dang2022} proposed a neural architecture to segment arteries and veins from different MR sequences. Artery segmentations from time-of-flight MRA volumes resulted in a Dice score coefficient of $79.32\%$, whereas vein segmentations were only assessed through visual inspection. A further limitation is that the network requires separate training for each modality.

\vspace{0.1cm}
\noindent
{\color{bmv@sectioncolor}\textbf{Domain Adaptation.}}
Supervised DA methods~\cite{Ghafoorian2017,Zhu2020} simplify model training by assuming that a small number of labeled data in the target domain are available. These methods, however, require labeled target data, which is particularly costly to obtain for brain vessels. Unsupervised DA (UDA) techniques avoid the use of target domain labels, leveraging the potential information available in readily accessible unlabeled data. These techniques can perform adaptation at the input/image-\cite{Yao2022}, feature-\cite{Wu2022} or output-\cite{SynthSeg} level; or a combination of two of the categories~\cite{SIFA}. Among these, image-level alignment methods have gained a lot of traction due to the success of image-to-image translation methods~\cite{CycleGAN}, which convert source-domain images to target-domain images. However, many image-alignment approaches rely on adversarial training~\cite{Ning2021,CycleGAN,SIFA}, which is known to be unstable. Lastly, only recently researchers have started to investigate semi-supervised domain adaptation for medical segmentation~\cite{Liu2022Semi,CS-CADA}.

Closer to our objective, two recent works rely on DA techniques to tackle domain shift for vessel segmentation. Peng et al.~\cite{DCDA} use a disentangled representation and two segmentation models, each specialized to its own domain, in a collaborative UDA learning module for 2D retinal segmentation. Gu et al.~\cite{CS-CADA} propose a semi-supervised method for cross-anatomy domain adaptation on 2D images that employs domain-specific batch normalization and cross-domain contrastive learning within a self-ensembling mean-teacher framework. Although these two approaches have reported promising results, we argue that the task of segmenting the cerebrovascular tree is more arduous, primarily due to the significant gap that exists between arteries and veins, as well as the intricate three-dimensional complexities inherent in their structure.

\section{Method}
\label{Method}
\begin{figure}[t]
\centering
\includegraphics[width=.8\textwidth]{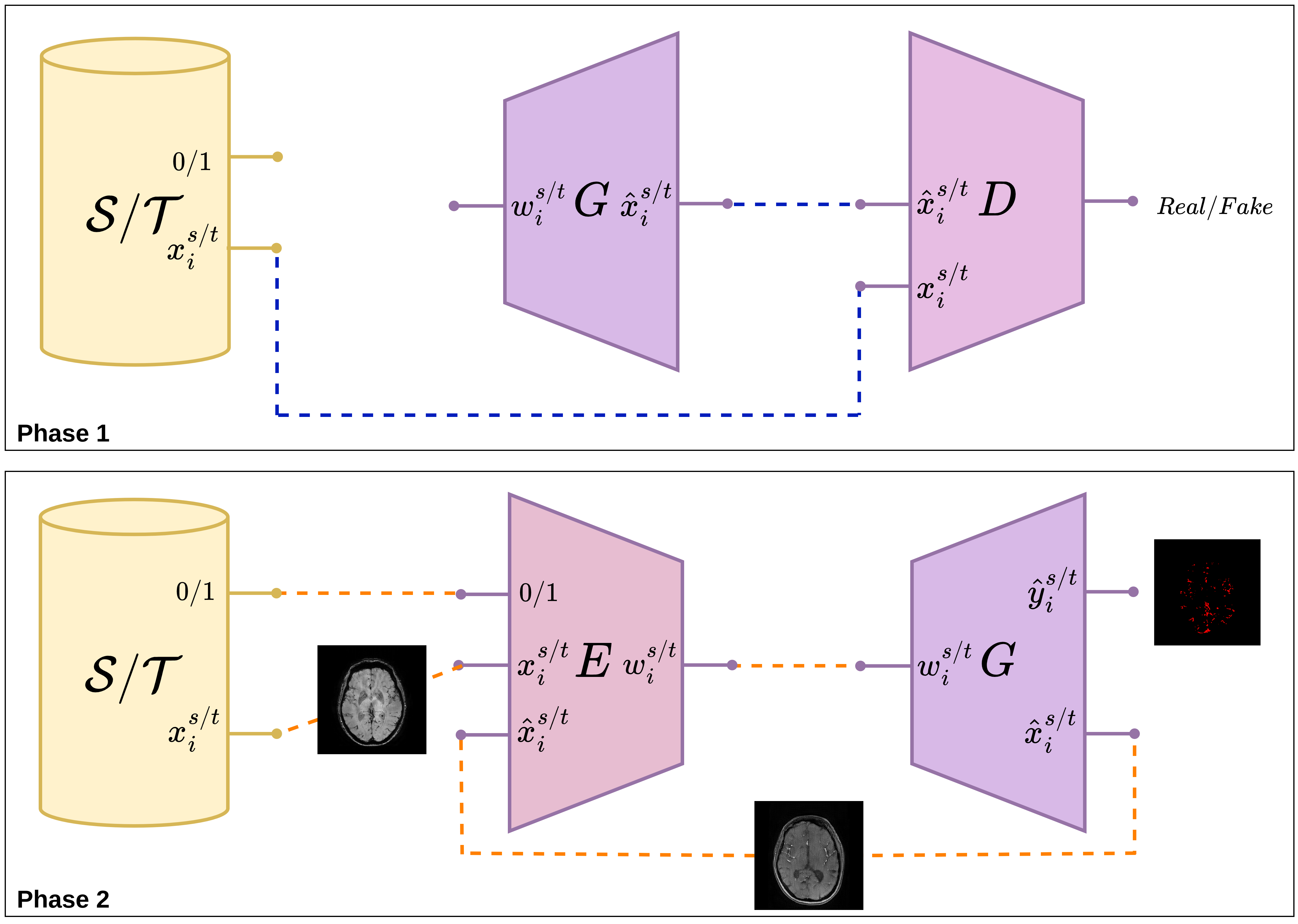}
\caption{Two-phase training algorithm. Images $x_i$ from domains $\mathcal{S}$ and $\mathcal{T}$ are fed into the model, composed of the generator $G$, the discriminator $D$ and the encoder $E$. The modules are trained in two separate phases. Phase 1 (top) trains $G$ in an adversarial fashion to learn a smooth and semantically rich latent space. Phase 2 (bottom) trains $E$ to perform image-to-image translation and refines $G$ to also generate segmentation masks $\hat{y}_i^t$ and $\hat{y}_i^s$. 
} \label{Fig:sketch}
\end{figure}

Let us denote $\mathcal{S}$ and $\mathcal{T}$ as the source and target domains. We denote $S = \{x_i^s, y_i^s\}_{i=1}^N$ a set of $N$ labeled images from $\mathcal{S}$, i.e., a set of labeled angiographies, $T_{U} = \{x_j^t\}_{j=1}^M$ a set of $M$ unlabeled images from $\mathcal{T}$, i.e., a set of unlabeled venographies, and $T_{L} = \{x_k^{lt}, y_k^{lt}\}_{k=1}^m$ a target labeled dataset consisting of $m \ll M$ annotated images, also from $\mathcal{T}$. Figure~\ref{Fig:sketch} illustrates our proposed end-to-end framework to segment venographies using annotated angiographic data in a semi-supervised fashion by encapsulating conditional image generation, image-to-image translation, and image segmentation.

Our framework consists of three elements: 1) a generator $G$, which is responsible for constructing a smooth and semantically rich latent space that can represent a heterogeneous range of images from $\mathcal{S}$ and $\mathcal{T}$ and for producing the final segmentations in $\mathcal{T}$; 2) a discriminator $D$ that enables the training of $G$; and 3) an encoder $E$ that performs image-to-image translation across domains. The three modules are trained in two phases. The first phase trains $G$ to build its smooth and semantically rich latent space $\mathcal{W}$. The second phase trains $E$ to achieve image-to-image translation. By decoupling the training into two phases, we guarantee that only the first phase involves an adversarial training routine, a highly desirable property to enforce stable and fast convergence.

\vspace{0.2cm}
\noindent
{\color{bmv@sectioncolor}\textbf{Phase 1.}}
$G$ is trained to learn a function $g$ that maps a latent vector $w_i \in \mathcal{W}$ into the corresponding image $\hat{x}_i$, in either $\mathcal{S}$ or $\mathcal{T}$. The images generated by $G$ are fed into the discriminator $D$, which is trained in an adversarial manner along with the generator $G$ to distinguish between real and fake samples. In the meantime, the generator's objective is to fool the discriminator by retrieving images of indistinguishable quality from the original images, for both $\mathcal{S}$ and $\mathcal{T}$.

Without loss of generality, we adopt a StyleGAN2 architecture~\cite{StyleGAN2} for $G$ and $D$, given its well-established ability to disentangle high-level features in the latent space $\mathcal{W}$ during training. The property of disentanglement grants the generator the ability to control and vary specific features of the generated images independently of others. This provides control over the generated images. Additionally, manipulation of features such as intensities, textures, location and shape of vessels can be handled separately, allowing the next phase to establish mappings between arteries and veins at different semantic levels.

\vspace{0.2cm}
\noindent
{\color{bmv@sectioncolor}\textbf{Phase 2.}}
The encoder $E$ is trained to learn a function $e$ that inverts the generative function $g$, i.e., discover the corresponding latent representation of an image in both $\mathcal{S}$ and $\mathcal{T}$. The main goal of $E$ is to perform image-to-image translation under the assumption that all vessel properties are disentangled within the latent space $\mathcal{W}$. Due to this property, the encoder is able to establish correspondences between the two source and target domains at a high level of abstraction~\cite{TransGaGa,Yang2022}, discovering mappings that differentiate between various vessel characteristics, such as altering vessel intensities, while preserving their spatial arrangement. To achieve this, $E$ is trained in three possible configurations, which are alternated by providing the model with an interchangeable binary label. When the label aligns with the domain of the inputted images, the encoder retrieves their reconstructions within the same domain (Configuration 2.1). When the label is inverted, the encoder retrieves inter-domain translations in the opposite domain, i.e., angiogram-to-venogram (Configuration 2.2) or venogram-to-angiogram (Configuration 2.3) translations. By feeding the model with opposite labels in succession, the encoder gains a cyclical behavior. This enables the computation of cycle-consistency losses without the need for two encoder-decoder pairs, which represents a reduction in complexity and computational cost compared to traditional cycle-based approaches~\cite{CycleGAN}.

\begin{figure}[ht]
\centering
\includegraphics[width=0.99\textwidth]{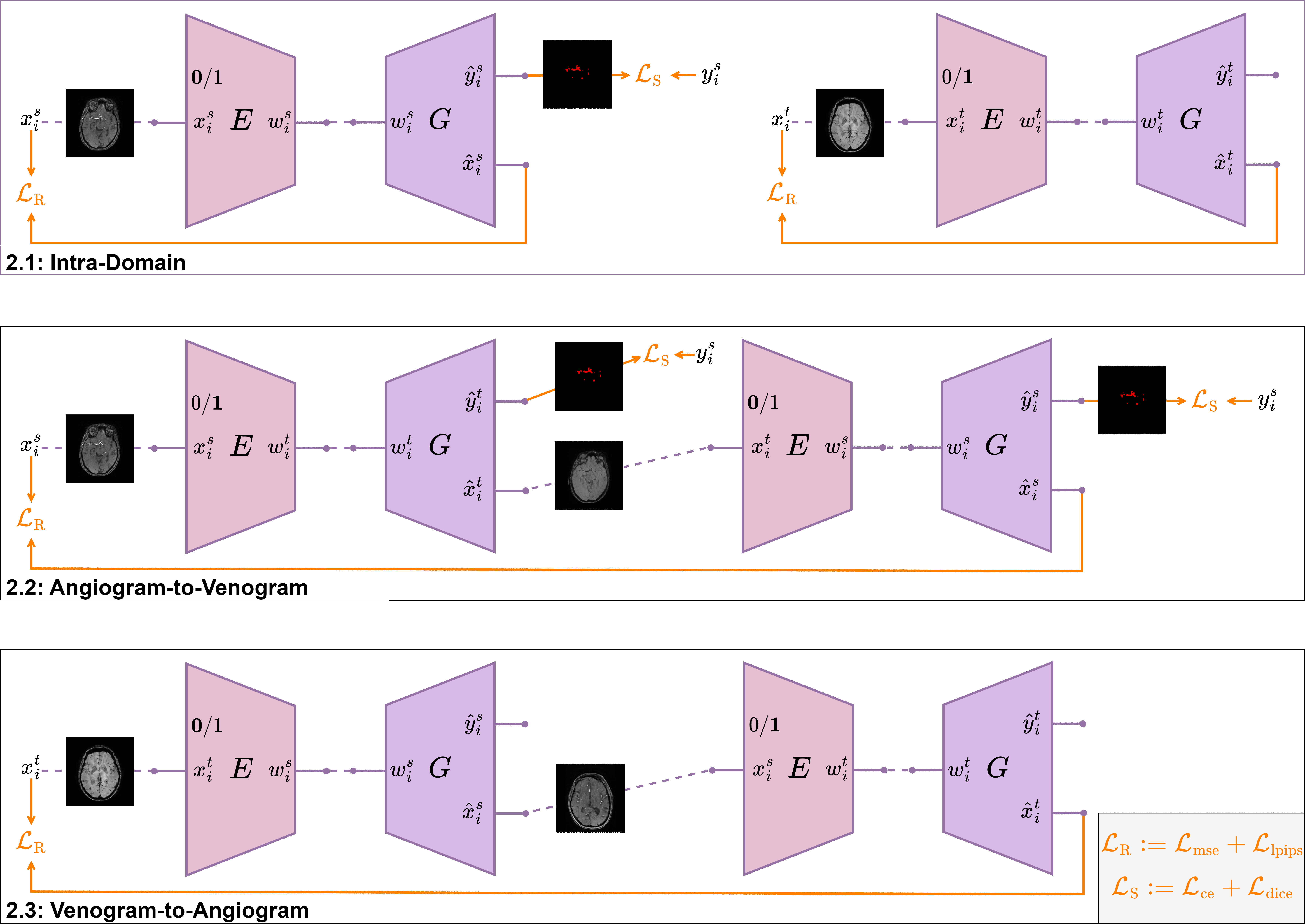}
\caption{
Phase 2 of the training algorithm alternating intra-domain (2.1) and inter-domain (2.2 and 2.3) configurations. 
We compute the sum of mean squared error and LPIPS~\cite{LPIPS} for reconstruction and cycle-consistency losses~$\mathcal{L}_{\textrm{\textsc{r}}}$, and the sum of Dice and cross-entropy for segmentation losses~$\mathcal{L}_{\textrm{\textsc{s}}}$. The backpropagation of $\mathcal{L}_{\textrm{\textsc{r}}}$ updates only the weights of $E$, while $\mathcal{L}_{\textrm{\textsc{s}}}$ affects both $E$ and $G$.
} \label{Fig:supplementary}
\end{figure}

In order to achieve segmentation, the generator is extended by adding a label-synthesis branch~\cite{DatasetGAN}, consisting of three fully-connected layers attached to the feature vectors of $G$. With this, we aim to avoid the use of a separate segmentation module, thus decreasing computational requirements. By freezing all other parameters in the framework, the branch is optimized in isolation, with the assistance of labeled samples from $S$ and $T_L$,  to return a semantic segmentation mask that aligns with a generated image. Differently from reconstruction and cycle-consistency, segmentation losses are computed only when labels are available, i.e., for $S$ and $T_L$. After one cycle, losses are propagated into $E$ and $G$ only once for the most recent pass. Figure~\ref{Fig:supplementary} provides a detailed breakdown of Phase 2. 




\vspace{0.2cm}
\noindent
{\color{bmv@sectioncolor}\textbf{Inference.}}
Given an unseen target image $x^t$, our model generates its reconstruction in the original target domain, i.e., $\hat{x}^t$, and its inter-domain translation, which accounts for translating $x^t$ into $\mathcal{S}$, i.e., $\hat{x}^s$. Additionally, $G$ generates the corresponding segmentation mask, $\hat{y}^t$ and $\hat{y}^s$, from its label-synthesis branch for both $\hat{x}^t$ and $\hat{x}^s$. Both masks contain valuable information pertaining to vessel segmentation within the target domain. For this reason, the final predicted segmentation mask is obtained as the average of $\hat{y}^t$ and $\hat{y}^s$ before the final argmax operation.

\section{Experiments and Results}
\label{Experiments}

\subsection{Experimental Setup}
{\color{bmv@sectioncolor}\textbf{Datasets and Preprocessing.}}
We use a subset of 49 time-of-flight (TOF) MRA volumes selected at random from the OASIS-3 database~\cite{LaMontagne2019}, with median dimension $576 \times 768 \times 232$ and voxel size $0.60 \times 0.30 \times 0.30$, as the source database $\mathcal{S}$. This selection includes 27 cognitively normal patients, as well as 10 patients at various stages of cognitive decline, all adults ranging in age from 42 to 95 years. 
For the target database $\mathcal{T}$, we use a set of 28 susceptibility weighted imaging (SWI) venographies, with median dimension $480 \times 480 \times 288$ and voxel size $0.50 \times 0.50 \times 0.50$.  The database was obtained from retrospective studies previously conducted at UCL Queen Square Institute of Neurology, Queen Square MS Centre, University College London and includes adult subjects with no lesions visible on SWI. 
The source database and 11 volumes in the target database have associated brain and vessel masks. Brain masks were obtained using SynthStrip~\cite{HOOPES2022119474} for the source database and HD-BET~\cite{Isensee2019} for the target database, while vessel masks were manually annotated by an expert.

Each volume was standardized based on its mean and standard deviation and rescaled to have a uniform spacing of [0.5, 0.5, 0.5] mm between adjacent voxels. Next, 2D slices were extracted from the volumes, and each slice was clipped and normalized in the range $[-1,+1]$. The slices were also padded and cropped to a fixed size of $512 \times 512$ pixels. The segmentation masks were one-hot encoded into a single label of size $3 \times 512 \times 512$ (with an extra dimension for the background).

\vspace{0.1cm}
\noindent
{\color{bmv@sectioncolor}\textbf{Setup.}} The training set consists of 45 TOF images (${S}$), 17 SWI venographies without annotations ($T_U$) and 3 annotated SWI venographies (${T}_L$). Four annotated SWIs are used for the validation set, whereas the remaining four annotated angiographies and four annotated venographies are kept for testing.

We compare our method against four DA state-of-the-art methods: \\
 \textbf{1)  Synergistic image and feature adaptation (SIFA)}~\cite{SIFA}, an UDA technique based on image-to-image translation for multi-class medical segmentation; \\
 \textbf{2) SynthSeg}~\cite{SynthSeg}, an UDA 3D output-level alignment method based on synthetic data generation for brain synthesis and segmentation; \\
 \textbf{3) Contrastive Semi-supervised learning for Cross Anatomy Domain Adaptation (CS-CADA)}~\cite{CS-CADA}, a semi-supervised DA method relying on feature alignment, which has been tested for 2D coronary artery segmentation; and \\
 \textbf{4) DCDA} \cite{DCDA}, an UDA technique based on style representation transfer, specifically designed for retinal vessel segmentation. 
 
Additionally, we include a Sato filter~\cite{sato1997} for vessel enhancement as a baseline model. The segmentation results obtained on the test set are quantitatively assessed using the Dice coefficient (Dice), the centerlineDice (clDice)~\cite{Shit2021}, precision and recall.

\vspace{0.1cm}
\noindent
{\color{bmv@sectioncolor}\textbf{Implementation Details.}} We implemented our framework in PyTorch 1.9.1. Both Phase 1 and Phase 2 were executed with a batch size of 4 images and trained respectively for 250k and 50k iterations. We retained the model with the best validation performance for the final evaluation. The generator $G$ and discriminator $D$ are based on the architectures of StyleGAN2~\cite{StyleGAN2}, while the label-synthesis branch is adapted from DatasetGAN~\cite{DatasetGAN}. 
The input images are encoded into the extended latent space $\mathcal{W}+$ of StyleGAN2, as formulated in~\cite{pixel2style2pixel}. The architecture of $E$ is inspired by~\cite{Residuals}, which utilizes a ResNet backbone and multiple output branches: one for latent code prediction, the others for feature tensor prediction. These are filtered and passed to $G$ through a dynamic skip connection module~\cite{Yang2022}, which establishes fine-level content correspondences.

For SIFA, SynthSeg, CS-CADA and DCDA, we used the implementations proposed in the original publications, following the respective training schedules. The Sato filter was implemented in C++ using the ITK library\footnote{<annonymized>}. We trained, validated and tested our proposed method as well as the state-of-the-art methods on two NVIDIA GeForce RTX 2080 Ti GPUs.

\subsection{Results}
We first assess the intra-domain vessel segmentation performance of our method using the angiography images within the test set. Our model achieves Dice of $79.3\pm4.4\%$, a clDice of $78.7\pm5.3\%$, precision of $83.1\pm1.4\%$ and a recall of 76.2$\pm7.4\%$. The obtained results are comparable with state-of-the-art methods for artery segmentation~\cite{livne2019,Dang2022}.

We evaluate our method's performance in the target domain, i.e., venographies, and we compare it against state-of-the-art DA methods and the Sato filter. For the sake of fairness, in addition to vessel segmentation, we report the results obtained for cross-modality brain segmentation since most methods (i.e., SIFA, SynthSeg and CS-CADA) have been originally conceived to segment large objects, such as the brain. Table~\ref{Tab:DSC} reports the obtained results.

While the brain segmentation results are generally satisfactory, most of the methods show poor performance at segmenting vessels, being surpassed by the simpler Sato filter. Despite being provided with only 3 target annotations, the proposed method succeeds in segmenting vessels by linking arteries and veins across the two modalities, achieving high performance in both domains.
In particular, our proposed method bridges the large domain gap that encompasses not only low-level features, such as intensities and textures, but also high-level features, such as variations in the locations and shapes of arteries and veins. 

\begin{figure}
\includegraphics[width=\textwidth]{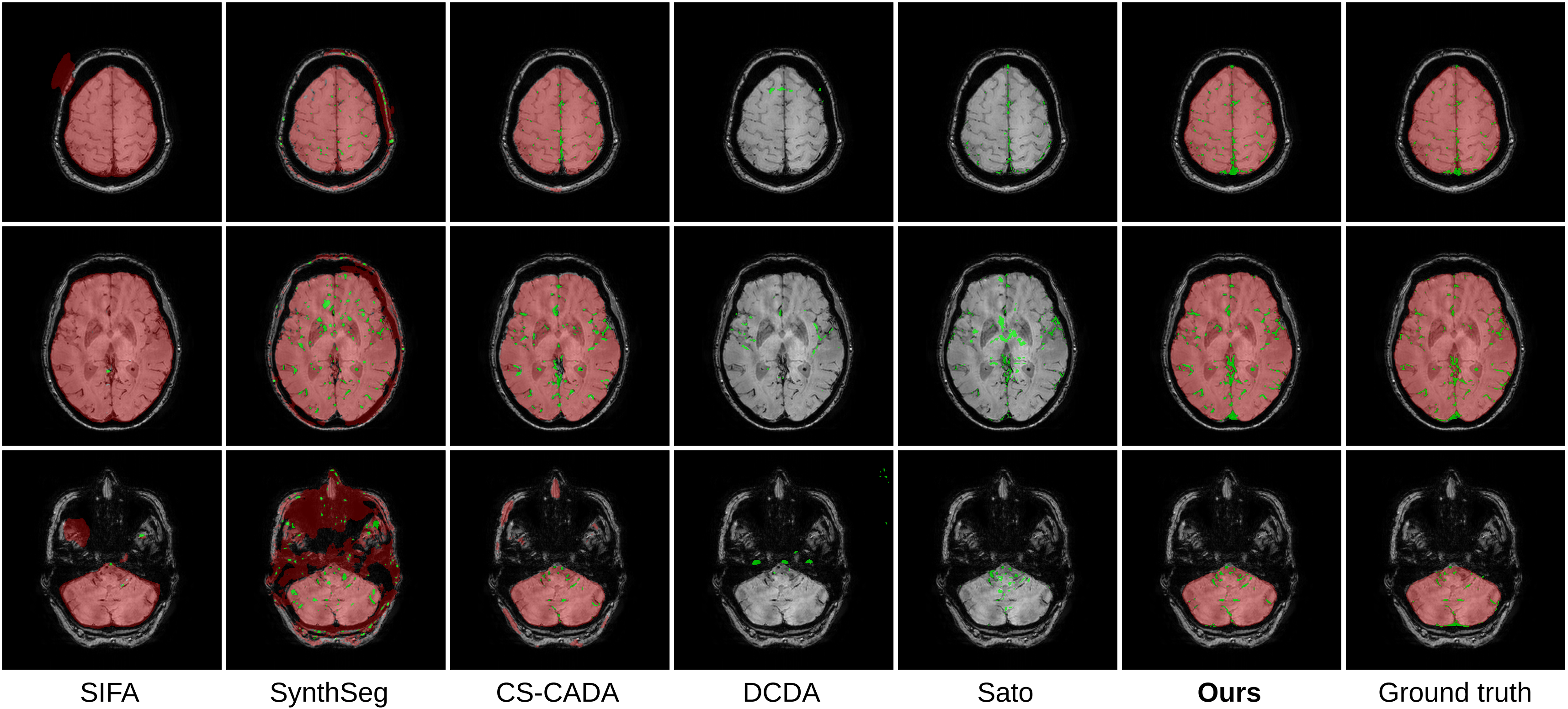}
\caption{Visual comparison of the results produced by different methods for brain and vein segmentation from SWI images. Brain masks are indicated in red, vessels in green. From the 1st to the 3rd row we display in order a top-level, a middle-level and a bottom-level slice.} \label{Fig:results}
\end{figure}

Figure~\ref{Fig:results} presents a visual comparison of the results at different locations of the brain. 
We observe that, in some cases, spatial correspondence is lost during translation, e.g., DCDA mixes top and bottom slices. Methods like SIFA focus on translating the overall appearance of the image, perhaps because it has been designed to deal with the adaptation of objects at a larger scale. Although CS-CADA achieves the second highest Dice on the brain, its performance drops when detecting veins in our testing set. We hypothesize that, despite being designed to address significant domain gaps, CS-CADA may not be well-suited for highly complex 3D problems. Our method is able to disentangle the volume-related image properties, such as spatial information and appearance, and the vessel-related properties, such as intensities, textures, shapes, and locations. As a consequence, our method generates a translation that resembles a TOF image in terms of style while retaining the SWI's content, thus leading to a fair vessel segmentation result.

\begin{table}[t]
\centering
\caption{Performance comparison of different DA methods in the target domain, i.e., four testing venographies. We report mean Dice, Precision, Recall and clDice (in \%) with standard deviations.}\label{Tab:DSC}
\renewcommand{\arraystretch}{1.2}%
\footnotesize
\begin{tabular}{llcccccc}
\bottomrule
 && SIFA & SynthSeg & CS-CADA & DCDA  & Sato & Ours\\
\hline
\multirow{ 2}{*}{Dice} & Vessels & 0.8 ± 0.2 & 37.3 ± 4.4 & 51.4 ± 1.7 & 4.5 ± 0.4 & 44.2 ± 7.2 & $\mathbf{70.4}$ ± $\mathbf{2.4}$\\
& Brain & 91.5 ± 0.4 & 79.6 ± 3.8 & 91.5 ± 0.8 & - & - & $\mathbf{97.5}$ ± $\mathbf{0.2}$\\
\hline
\multirow{ 2}{*}{Precision} & Vessels & 11.6 ± 1.2 & 42.3 ± 9.2 & 58.6 ± 6.7 & 14.8 ± 3.5 & 42.7 ± 6.4 & $\mathbf{66.8}$ ± $\mathbf{5.2}$\\
& Brain & 84.8 ± 0.7 & 69.4 ± 5.5 & 89.6 ± 0.8 & - & - & $\mathbf{97.6}$ ± $\mathbf{0.3}$\\
\hline
\multirow{ 2}{*}{Recall} & Vessels & 0.4 ± 0.1 & 33.9 ± 1.6 & 46.2 ± 2.2 & 2.7 ± 0.2 & 46.1 ± 9.3 & $\mathbf{74.9}$ ± $\mathbf{3.0}$\\
& Brain & $\mathbf{99.3}$ ± $\mathbf{0.1}$ & 93.6 ± 0.5 & 93.5 ± 1.1 & - & - & 97.4 ± 0.5\\
\hline
clDice & Vessels & 0.8 ± 0.2 & 48.2 ± 4.7 & 58.0 ± 2.8 & 3.9 ± 0.2 & 50.0 ± 6.7 & $\mathbf{74.8}$ ± $\mathbf{2.4}$\\
\toprule
\end{tabular}
\end{table}

\section{Discussion and Conclusion}
\label{Conclusion}

In this paper, we proposed an end-to-end semi-supervised domain adaptation framework for accurately segmenting 3D brain vessels using primarily annotations from arterial images, which are easier to obtain. We addressed the challenge of domain shift caused by the differences in imaging modalities, not only concerning the general appearance but also the different vessels they target (arteries or veins). Our approach relies on the StyleGAN2 architecture, which allows to represent heterogeneous volumetric data and bridge the large domain gap between angiography and venography brain images.

In accordance with the goal of reducing complexity in cycle-based architectures, our network was optimized and constructed from a single generator, discriminator, and encoder. We highlight how this design brings an advantage in terms of memory requirements and model comprehensibility, facilitating its debugging and understanding during experimentation and its deployment in real-world scenarios. Our method performs conventional image-level alignment with high performance, despite other techniques, such as SIFA or DCDA, build on more complex architectures. In particular, SIFA counts three distinct discriminators, two encoder-decoder networks and one segmentation module, while DCDA includes four encoder-generator pairs and two segmentation networks. These differences demonstrate the efficacy of our approach in building an efficient and intuitive model without sacrificing performance. Moreover, its training was stabilized and accelerated by designing a two-phase learning algorithm that minimizes the use of adversarial training.

Although our approach outperforms DA state-of-the-art methods, we acknowledge that there is still room for improvement, and this problem remains a challenging task. To the best of our knowledge, this is the first attempt to achieve artery-to-vein translation. Nevertheless, our results show a promising performance at segmenting the brain veins in a semi-supervised domain adaptation scenario, despite the inherent difficulty of the problem posed by the significant gap between veins and arteries and the intricate morphology of the cerebrovascular tree. Overall, our work highlights the potential of domain adaptation methodologies for improving brain vessel segmentation, and we hope it can inspire further research in this area.

\section*{Acknowledgments}
FG, NB and MAZ are partially funded by the French government via Investments in the Future project (3IA Côte d’Azur, PRAIRIE 3IA Institute) managed by the ANR (ANR-19-P3IA-000, ANR-19-P3IA-0001). DF and MAZ are funded by the ANR JCJC project I-VESSEG (22-CE45-0015-01). NB has received funding from the ANR-10-IA Institut Hospitalo-Universitaire-6 (ANR-10-IAIHU-06).
\bibliography{references}

\begin{thebibliography}{33}
\providecommand{\natexlab}[1]{#1}
\providecommand{\url}[1]{\texttt{#1}}
\expandafter\ifx\csname urlstyle\endcsname\relax
  \providecommand{\doi}[1]{doi: #1}\else
  \providecommand{\doi}{doi: \begingroup \urlstyle{rm}\Url}\fi

\bibitem[Billot et~al.(2023)Billot, Greve, Puonti, Thielscher, Van~Leemput,
  Fischl, Dalca, Iglesias, et~al.]{SynthSeg}
Benjamin Billot, Douglas~N Greve, Oula Puonti, Axel Thielscher, Koen
  Van~Leemput, Bruce Fischl, Adrian~V Dalca, Juan~Eugenio Iglesias, et~al.
\newblock Synth{S}eg: Segmentation of brain {MRI} scans of any contrast and
  resolution without retraining.
\newblock \emph{Medical Image Analysis}, page 102789, 2023.

\bibitem[Chen et~al.(2019)Chen, Dou, Chen, Qin, and Heng]{SIFA}
Cheng Chen, Qi~Dou, Hao Chen, Jing Qin, and Pheng-Ann Heng.
\newblock Synergistic image and feature adaptation: Towards cross-modality
  domain adaptation for medical image segmentation.
\newblock In \emph{Proceedings of the AAAI Conference on Artificial
  Intelligence}, pages 865--872, 2019.

\bibitem[Dang et~al.(2022)Dang, Galati, Cortese, {Di Giacomo}, Marconetto,
  Mathur, Lekadir, Lorenzi, Prados, and Zuluaga]{Dang2022}
Vien~Ngoc Dang, Francesco Galati, Rosa Cortese, Giuseppe {Di Giacomo}, Viola
  Marconetto, Prateek Mathur, Karim Lekadir, Marco Lorenzi, Ferran Prados, and
  Maria~A. Zuluaga.
\newblock {Vessel-CAPTCHA}: An efficient learning framework for vessel
  annotation and segmentation.
\newblock \emph{Medical Image Analysis}, 75:\penalty0 102263, 2022.

\bibitem[Ghafoorian et~al.(2017)Ghafoorian, Mehrtash, Kapur, Karssemeijer,
  Marchiori, Pesteie, Guttmann, de~Leeuw, Tempany, van Ginneken, Fedorov,
  Abolmaesumi, Platel, and Wells]{Ghafoorian2017}
Mohsen Ghafoorian, Alireza Mehrtash, Tina Kapur, Nico Karssemeijer, Elena
  Marchiori, Mehran Pesteie, Charles R.~G. Guttmann, Frank-Erik de~Leeuw,
  Clare~M. Tempany, Bram van Ginneken, Andriy Fedorov, Purang Abolmaesumi, Bram
  Platel, and William~M. Wells.
\newblock Transfer learning for domain adaptation in {MRI}: Application in
  brain lesion segmentation.
\newblock In \emph{Medical Image Computing and Computer Assisted Intervention
  -- {MICCAI}}, pages 516--524, 2017.

\bibitem[Gu et~al.(2023)Gu, Zhang, Wang, Lei, Song, Zhang, Li, and
  Zhang]{CS-CADA}
Ran Gu, Jingyang Zhang, Guotai Wang, Wenhui Lei, Tao Song, Xiaofan Zhang, Kang
  Li, and Shaoting Zhang.
\newblock Contrastive semi-supervised learning for domain adaptive segmentation
  across similar anatomical structures.
\newblock \emph{IEEE Transactions on Medical Imaging}, 42\penalty0
  (1):\penalty0 245--256, 2023.

\bibitem[Guan and Liu(2022)]{Guan2022}
Hao Guan and Mingxia Liu.
\newblock Domain adaptation for medical image analysis: A survey.
\newblock \emph{IEEE Transactions on Biomedical Engineering}, 69\penalty0
  (3):\penalty0 1173--1185, 2022.

\bibitem[Helthuis et~al.(2019)Helthuis, Van~Doormaal, Hillen, Bleys, Harteveld,
  Hendrikse, Van~der Toorn, Brozici, Zwanenburg, and Van~der
  Zwan]{Helthuis2019}
Jasper~HG Helthuis, Tristan~PC Van~Doormaal, Berend Hillen, Ronald~LAW Bleys,
  Anita~A Harteveld, Jeroen Hendrikse, Annette Van~der Toorn, Mariana Brozici,
  Jaco~JM Zwanenburg, and Albert Van~der Zwan.
\newblock Branching pattern of the cerebral arterial tree.
\newblock \emph{The Anatomical Record}, 302\penalty0 (8):\penalty0 1434--1446,
  2019.

\bibitem[Hoopes et~al.(2022)Hoopes, Mora, Dalca, Fischl, and
  Hoffmann]{HOOPES2022119474}
Andrew Hoopes, Jocelyn~S. Mora, Adrian~V. Dalca, Bruce Fischl, and Malte
  Hoffmann.
\newblock {SynthStrip:} skull-stripping for any brain image.
\newblock \emph{NeuroImage}, 260:\penalty0 119474, 2022.

\bibitem[Isensee et~al.(2019)Isensee, Schell, Pflueger, Brugnara, Bonekamp,
  Neuberger, Wick, Schlemmer, Heiland, Wick, Bendszus, Maier-Hein, and
  Kickingereder]{Isensee2019}
Fabian Isensee, Marianne Schell, Irada Pflueger, Gianluca Brugnara, David
  Bonekamp, Ulf Neuberger, Antje Wick, Heinz-Peter Schlemmer, Sabine Heiland,
  Wolfgang Wick, Martin Bendszus, Klaus~H. Maier-Hein, and Philipp
  Kickingereder.
\newblock Automated brain extraction of multisequence {MRI} using artificial
  neural networks.
\newblock \emph{Human Brain Mapping}, 40\penalty0 (17):\penalty0 4952--4964,
  2019.

\bibitem[Kamnitsas et~al.(2017)Kamnitsas, Baumgartner, Ledig, Newcombe,
  Simpson, Kane, Menon, Nori, Criminisi, Rueckert, et~al.]{Kamnitsas2017a}
Konstantinos Kamnitsas, Christian Baumgartner, Christian Ledig, Virginia
  Newcombe, Joanna Simpson, Andrew Kane, David Menon, Aditya Nori, Antonio
  Criminisi, Daniel Rueckert, et~al.
\newblock Unsupervised domain adaptation in brain lesion segmentation with
  adversarial networks.
\newblock In \emph{Information Processing in Medical Imaging (IPMI)}, pages
  597--609, 2017.

\bibitem[Karras et~al.(2020)Karras, Laine, Aittala, Hellsten, Lehtinen, and
  Aila]{StyleGAN2}
Tero Karras, Samuli Laine, Miika Aittala, Janne Hellsten, Jaakko Lehtinen, and
  Timo Aila.
\newblock Analyzing and improving the image quality of stylegan.
\newblock In \emph{Proceedings of the IEEE/CVF Conference on Computer Vision
  and Pattern Recognition (CVPR)}, pages 8110--8119, 2020.

\bibitem[LaMontagne et~al.(2019)LaMontagne, Benzinger, Morris, Keefe, Hornbeck,
  Xiong, Grant, Hassenstab, Moulder, Vlassenko, Raichle, Cruchaga, and
  Marcus]{LaMontagne2019}
Pamela~J. LaMontagne, Tammie~LS. Benzinger, John~C. Morris, Sarah Keefe, Russ
  Hornbeck, Chengjie Xiong, Elizabeth Grant, Jason Hassenstab, Krista Moulder,
  Andrei~G. Vlassenko, Marcus~E. Raichle, Carlos Cruchaga, and Daniel Marcus.
\newblock {OASIS}-3: Longitudinal neuroimaging, clinical, and cognitive dataset
  for normal aging and alzheimer disease.
\newblock \emph{medRxiv}, 2019.

\bibitem[Lin et~al.(2018)Lin, Rawal, Agid, and Mandell]{Lin2018}
Amy Lin, Sapna Rawal, Ronit Agid, and Daniel~M Mandell.
\newblock Cerebrovascular imaging: which test is best?
\newblock \emph{Neurosurgery}, 83\penalty0 (1):\penalty0 5--18, 2018.

\bibitem[Liu et~al.(2022)Liu, Xing, Shusharina, Lim, Jay~Kuo, El~Fakhri, and
  Woo]{Liu2022Semi}
Xiaofeng Liu, Fangxu Xing, Nadya Shusharina, Ruth Lim, C-C Jay~Kuo, Georges
  El~Fakhri, and Jonghye Woo.
\newblock {ACT}: Semi-supervised domain-adaptive medical image segmentation
  with asymmetric co-training.
\newblock In \emph{Medical Image Computing and Computer Assisted Intervention
  -- {MICCAI}}, pages 66--76, 2022.

\bibitem[Livne et~al.(2019)Livne, Rieger, Aydin, Taha, Akay, Kossen, Sobesky,
  Kelleher, Hildebrand, Frey, and Madai]{livne2019}
Michelle Livne, Jana Rieger, Orhun~Utku Aydin, Abdel~Aziz Taha, Ela~Marie Akay,
  Tabea Kossen, Jan Sobesky, John~D. Kelleher, Kristian Hildebrand, Dietmar
  Frey, and Vince~I. Madai.
\newblock A {U-Net} deep learning framework for high performance vessel
  segmentation in patients with cerebrovascular disease.
\newblock \emph{Frontiers in Neuroscience}, 13:\penalty0 97, 2019.

\bibitem[Moccia et~al.(2018)Moccia, De~Momi, El~Hadji, and Mattos]{Moccia2018}
Sara Moccia, Elena De~Momi, Sara El~Hadji, and Leonardo~S Mattos.
\newblock Blood vessel segmentation algorithms—review of methods, datasets
  and evaluation metrics.
\newblock \emph{Computer Methods and Programs in Biomedicine}, 158:\penalty0
  71--91, 2018.

\bibitem[Ning et~al.(2021)Ning, Bian, Wei, Yu, Yuan, Wang, Guo, Ma, and
  Zheng]{Ning2021}
Munan Ning, Cheng Bian, Dong Wei, Shuang Yu, Chenglang Yuan, Yaohua Wang, Yang
  Guo, Kai Ma, and Yefeng Zheng.
\newblock A new bidirectional unsupervised domain adaptation segmentation
  framework.
\newblock In \emph{Information Processing in Medical Imaging (IPMI)}, pages
  492--503, 2021.

\bibitem[Passat et~al.(2007)Passat, Ronse, Baruthio, Armspach, and
  Foucher]{passat2007}
Nicolas Passat, Christian Ronse, Joseph Baruthio, J-P Armspach, and Jack
  Foucher.
\newblock Watershed and multimodal data for brain vessel segmentation:
  Application to the superior sagittal sinus.
\newblock \emph{Image and Vision Computing}, 25\penalty0 (4):\penalty0
  512--521, 2007.

\bibitem[Peng et~al.(2022)Peng, Lin, Cheng, Huang, and Tang]{DCDA}
Linkai Peng, Li~Lin, Pujin Cheng, Ziqi Huang, and Xiaoying Tang.
\newblock Unsupervised domain adaptation for cross-modality retinal vessel
  segmentation via disentangling representation style transfer and
  collaborative consistency learning.
\newblock In \emph{International Symposium on Biomedical Imaging (ISBI)}, pages
  1--5, 2022.

\bibitem[Richardson et~al.(2021)Richardson, Alaluf, Patashnik, Nitzan, Azar,
  Shapiro, and Cohen-Or]{pixel2style2pixel}
Elad Richardson, Yuval Alaluf, Or~Patashnik, Yotam Nitzan, Yaniv Azar, Stav
  Shapiro, and Daniel Cohen-Or.
\newblock Encoding in style: A stylegan encoder for image-to-image translation.
\newblock In \emph{Proceedings of the IEEE/CVF Conference on Computer Vision
  and Pattern Recognition (CVPR)}, pages 2287--2296, 2021.

\bibitem[Sato et~al.(1997)Sato, Nakajima, Atsumi, Koller, Gerig, Yoshida, and
  Kikinis]{sato1997}
Yoshinobu Sato, Shin Nakajima, Hideki Atsumi, Thomas Koller, Guido Gerig,
  Shigeyuki Yoshida, and Ron Kikinis.
\newblock {3D multi-scale line filter for segmentation and visualization of
  curvilinear structures in medical images}.
\newblock In \emph{CVRMed-MRCAS: First Joint Conference Computer Vision,
  Virtual Reality and Robotics in Medicine and Medical Robotics and
  Computer-Assisted Surgery}, pages 213--222, 1997.

\bibitem[Shit et~al.(2021)Shit, Paetzold, Sekuboyina, Ezhov, Unger, Zhylka,
  Pluim, Bauer, and Menze]{Shit2021}
Suprosanna Shit, Johannes~C. Paetzold, Anjany Sekuboyina, Ivan Ezhov, Alexander
  Unger, Andrey Zhylka, Josien P.~W. Pluim, Ulrich Bauer, and Bjoern~H. Menze.
\newblock cldice - a novel topology-preserving loss function for tubular
  structure segmentation.
\newblock In \emph{Proceedings of the IEEE/CVF Conference on Computer Vision
  and Pattern Recognition (CVPR)}, pages 16560--16569, 2021.

\bibitem[Tsai et~al.(2018)Tsai, Hung, Schulter, Sohn, Yang, and
  Chandraker]{Tsai2018}
Yi-Hsuan Tsai, Wei-Chih Hung, Samuel Schulter, Kihyuk Sohn, Ming-Hsuan Yang,
  and Manmohan Chandraker.
\newblock Learning to adapt structured output space for semantic segmentation.
\newblock In \emph{Proceedings of the IEEE Conference on Computer Vision and
  Pattern Recognition (CVPR)}, pages 7472--7481, 2018.

\bibitem[Wu et~al.(2022)Wu, Gu, Dong, Wang, and Zhang]{Wu2022}
Jianghao Wu, Ran Gu, Guiming Dong, Guotai Wang, and Shaoting Zhang.
\newblock {FPL-UDA}: Filtered pseudo label-based unsupervised cross-modality
  adaptation for vestibular schwannoma segmentation.
\newblock In \emph{International Symposium on Biomedical Imaging (ISBI)}, pages
  1--5, 2022.

\bibitem[Wu et~al.(2019)Wu, Cao, Li, Qian, and Loy]{TransGaGa}
Wayne Wu, Kaidi Cao, Cheng Li, Chen Qian, and Chen~Change Loy.
\newblock {TransGaGa}: Geometry-aware unsupervised image-to-image translation.
\newblock In \emph{Proceedings of the IEEE/CVF Conference on Computer Vision
  and Pattern Recognition (CVPR)}, June 2019.

\bibitem[Yang et~al.(2022)Yang, Jiang, Liu, and Loy]{Yang2022}
Shuai Yang, Liming Jiang, Ziwei Liu, and Chen~Change Loy.
\newblock Unsupervised image-to-image translation with generative prior.
\newblock In \emph{Proceedings of the IEEE/CVF Conference on Computer Vision
  and Pattern Recognition (CVPR)}, pages 18332--18341, 2022.

\bibitem[Yao et~al.(2022{\natexlab{a}})Yao, Su, Huang, Yang, Sun, Hussain, and
  Coenen]{Yao2022}
Kai Yao, Zixian Su, Kaizhu Huang, Xi~Yang, Jie Sun, Amir Hussain, and Frans
  Coenen.
\newblock A novel 3d unsupervised domain adaptation framework for
  cross-modality medical image segmentation.
\newblock \emph{IEEE Journal of Biomedical and Health Informatics}, 26\penalty0
  (10):\penalty0 4976--4986, 2022{\natexlab{a}}.

\bibitem[Yao et~al.(2022{\natexlab{b}})Yao, Newson, Gousseau, and
  Hellier]{Residuals}
Xu~Yao, Alasdair Newson, Yann Gousseau, and Pierre Hellier.
\newblock A style-based gan encoder for high fidelity reconstruction of images
  and videos.
\newblock In \emph{European Conference on Computer Vision - ECCV}, pages
  581--597, 2022{\natexlab{b}}.

\bibitem[Zhang et~al.(2018)Zhang, Isola, Efros, Shechtman, and Wang]{LPIPS}
Richard Zhang, Phillip Isola, Alexei~A. Efros, Eli Shechtman, and Oliver Wang.
\newblock The unreasonable effectiveness of deep features as a perceptual
  metric.
\newblock In \emph{Proceedings of the IEEE Conference on Computer Vision and
  Pattern Recognition (CVPR)}, pages 586--595, 2018.

\bibitem[Zhang et~al.(2021)Zhang, Ling, Gao, Yin, Lafleche, Barriuso, Torralba,
  and Fidler]{DatasetGAN}
Yuxuan Zhang, Huan Ling, Jun Gao, Kangxue Yin, Jean-Francois Lafleche, Adela
  Barriuso, Antonio Torralba, and Sanja Fidler.
\newblock Dataset{GAN}: Efficient labeled data factory with minimal human
  effort.
\newblock In \emph{Proceedings of the IEEE/CVF Conference on Computer Vision
  and Pattern Recognition (CVPR)}, pages 10145--10155, 2021.

\bibitem[Zhu et~al.(2017)Zhu, Park, Isola, and Efros]{CycleGAN}
Jun-Yan Zhu, Taesung Park, Phillip Isola, and Alexei~A. Efros.
\newblock Unpaired image-to-image translation using cycle-consistent
  adversarial networks.
\newblock In \emph{Proceedings of the IEEE International Conference on Computer
  Vision (ICCV)}, pages 2223--2232, 2017.

\bibitem[Zhu et~al.(2020)Zhu, Du, and Yan]{Zhu2020}
Qikui Zhu, Bo~Du, and Pingkun Yan.
\newblock Boundary-weighted domain adaptive neural network for prostate {MR}
  image segmentation.
\newblock \emph{IEEE Transactions on Medical Imaging}, 39\penalty0
  (3):\penalty0 753--763, 2020.

\bibitem[Zuluaga et~al.(2015)Zuluaga, Rodionov, Nowell, Achhala, Zombori,
  Mendelson, Cardoso, Miserocchi, McEvoy, Duncan, et~al.]{zuluaga2015}
Maria~A Zuluaga, Roman Rodionov, Mark Nowell, Sufyan Achhala, Gergely Zombori,
  Alex~F Mendelson, M~Jorge Cardoso, Anna Miserocchi, Andrew~W McEvoy, John~S
  Duncan, et~al.
\newblock Stability, structure and scale: improvements in multi-modal vessel
  extraction for seeg trajectory planning.
\newblock \emph{International Journal of Computer Assisted Radiology and
  Surgery}, 10:\penalty0 1227--1237, 2015.

\end{thebibliography}
\end{document}